\documentclass[pra,twocolumn,tightenlines,showpacs,nofootinbib]{revtex4}
\usepackage{multirow}
\usepackage{amsmath}
\usepackage{bm}
\usepackage{dcolumn}

\begin{document}
\title{Ytterbium in quantum gases and atomic clocks:
van der Waals interactions and blackbody shifts}

\author{M.~S.~Safronova$^1$}
\author{S.~G.~Porsev$^{1,2}$}
\author{Charles W. Clark$^3$}
\affiliation{ $^1$Department of Physics and Astronomy, University of Delaware,
    Newark, Delaware 19716, USA\\
$^2$Petersburg Nuclear Physics Institute, Gatchina, Leningrad District, 188300, Russia \\$^3$Joint
Quantum Institute, National Institute of Standards and Technology and the \\University of Maryland,
Gaithersburg, Maryland, 20899-8410, USA}
\date{\today}

\begin{abstract}
We evaluated the $C_6$ coefficients of Yb-Yb and Yb -alkali/group II  van der Waals interactions
with 2\% uncertainty. The only existing results for such quantities are for the Yb-Yb dimer. Our
value, $C_6=1929(39)$~a.u., is in excellent agreement with the recent experimental determination of
1932(35)~a.u. We have also developed a new  approach for the calculation of the dynamic correction
to the blackbody radiation shift. We have calculated this quantity for the Yb $6s^{2}~ ^1\!S_0 -
6s6p ~^3\!P_0$ clock transition  with 3.5\% uncertainty. This reduces the fractional uncertainty
due to the blackbody radiation shift in the Yb optical clock at 300 K  to $10^{-18}$ level.
\end{abstract}
 \maketitle

\section{Introduction}
Ytterbium (Yb: Z=70) has recently emerged as a subject of great interest in ultracold chemistry, physics, and metrology. For example, the first
state-resolved observation  of ultracold chemical reactions was recently reported for the Yb$^+$ + Rb$ \longrightarrow$ Yb + Rb$^+$
system~\cite{RatZipSia12}. Yb is a favorite candidate for the studies of ultracold gas mixtures. For example, Li and Yb mixtures have recently been
brought to simultaneous quantum degeneracy ~\cite{LiYb,LiYb1,LiYb3,LiYb4}. Controlled production of ultracold YbRb* molecules by photoassociation in
a mixure of Rb and Yb gases was recently reported in \cite{RbYb}.
 Such mixtures are of
interest for producing ultracold polar molecules for study of dipolar
quantum matter, fundamental symmetry studies, and many-body quantum
simulation ~\cite{LiYb4}.  The
 availability of 5 bosonic and 2 fermionic stable isotopes
makes Yb especially attractive for studies of multicomponent
superfluids.

 The spectrum of Yb contains a number of long-lived excited states
 that are conveniently accessed by optical techniques.
This makes Yb an excellent candidate for atomic parity violation
(APV) studies that  test
 the Standard Model of electroweak interactions, put limits on its
  possible extensions, constrain parameters of weak hadronic interactions,
  and may yield information on  neutron
 distributions within nuclei ~\cite{PNC,PNC2}.
  The APV signal recently observed in
  the Yb $6s^{2}~ ^1S_0  - 5d6s ~^3D_1$ 408-nm forbidden transition~\cite{PNC1,PNC2}
   is two orders of magnitude larger than in Cs, subject of the most
 accurate APV study to date. Such long-lived states are also convenient for the
development of next-generation ultra-precise frequency standards. The Yb $^{1}\!S_{0}\rightarrow \,
^{3}\!P_{0}^{o}$ 578-nm transition now provides one of the world's most accurate optical atomic
frequency standards~\cite{Ybclock,YbclockBBR}.

The work carried out in this Letter is pertinent to all applications mentioned above. Our two main
subjects are the determination of van der Waals $C_6$ coefficients that characterize the long-range
interactions  between two atoms, and the blackbody radiation (BBR) shifts of the two states in
atomic clock transitions. Knowledge of the long range interactions in Yb-Yb and Yb-alkali/group II
dimers is critical to understanding the physics of dilute gas mixtures. The
 dynamic correction to the BBR shift is one of the  largest
 irreducible contributions to
uncertainty budget of the Yb clock~\cite{Ybclock}, and it is
difficult to determine experimentally.
 These two seemingly disparate
topics both require accurate determination of frequency-dependent
atomic  polarizabilities over  a wide range of frequencies.
Therefore, it is natural to consider them in the same work. A future
accurate theoretical determination of the APV amplitude in Yb
requires a similar approach, and this work provides a background for
such studies.

We carry out the  calculation of frequency-dependent atomic polarizabilities  using the
first-principles approach that combines configuration interaction (CI) with the coupled-cluster
all-order approach (CI+all-order) that treats both core and valence correlation to all orders.
Several new method developments are presented in this work. First, we have implemented the reduced
linear equation (RLE) and direct inversion in iterative subspace (DIIS) stabilizer procedures
described in Ref.~\cite{RLE} into the coupled-cluster part of the CI+all-order method. Otherwise,
the construction of the effective Hamiltonian needed for the incorporation of the core and
core-valence correlations into the CI method could not be carried out due to convergence problems
associated with extremely large correlations involving the 4f shell. Second, we have applied the
CI+all-order method for the first time to the calculation of $C_6$ coefficients. Finally, we have
developed a new approach to the calculation of the dynamic correction to the blackbody radiation
(BBR) shift in terms of the second partial derivative  with respect to frequency of the dynamic
polarizability, as  obtained from the solution of the inhomogeneous equation in the valence sector.
Previous calculations of the dynamic correction to the BBR shift accounted for the contributions of
just a few intermediate states to the polarizability ~\cite{PorDer06}.

Unless stated otherwise, we use atomic units (a.u.) for all matrix
elements and polarizabilities throughout this paper: the numerical
values of the elementary
 charge, $e$, the reduced Planck constant, $\hbar = h/2
\pi$, and the electron mass, $m_e$, are set equal to 1. The atomic
unit for polarizability can be converted to SI units via
$\alpha/h$~[Hz/(V/m)$^2$]=2.48832$\times10^{-8}\alpha$~(a.u.), where
the conversion coefficient is $4\pi \epsilon_0 a^3_0/h$,
%where
$a_0$ is the Bohr radius and
$\epsilon_0$ is the electric constant.

Calculation of  Yb  properties requires an accurate treatment of both
core-valence and valence-valence correlations. This can be
accomplished within the framework of the CI+all-order method  that
combines configuration interaction and coupled-cluster approaches
\cite{SafKozJoh09,SafKozCla11,Tl}.
 Here we report the extension of
this method that resolves the convergence problems associated with particularly
 large correlation corrections as well as apply it for the first time to the calculation of the $C_6$ coefficients.
We refer the reader to Refs. ~\cite{SafKozJoh09,SafKozCla11} for detailed description of this
approach, and here we report only new method developments specific to this work. In order to
establish the accuracy of our approach, we also perform the pure CI and the CI combined with
many-body perturbation theory (CI+MBPT) calculations carried out with the same parameters such as
basis set, configuration space, number of partial waves, etc..

 The single-electron energies and the wave functions are found
from the solution of the Dirac-Hartree-Fock (DHF) equations. Then the wave functions and the
low-lying energy levels are determined by solving the multiparticle relativistic equation for two
valence electrons~\cite{KotTup87}, $ H_{\rm eff}(E_n) \Phi_n = E_n \Phi_n. $ The effective
Hamiltonian is defined as $ H_{\rm eff}(E) = H_{\rm FC} + \Sigma(E), $ where $H_{\rm FC}$ is the
Hamiltonian in the frozen-core approximation. The energy-dependent operator $\Sigma(E)$ which takes
into account virtual core excitations is constructed using second-order perturbation theory in the
CI+MBPT method \cite{DzuFlaKoz96b} and using a linearized coupled-cluster single-double method in
the CI+all-order approach \cite{SafKozJoh09}. However,  the CI+all-order approach developed in
\cite{SafKozJoh09,SafKozCla11} could not be directly implemented for Yb owing to convergence
problems of the all-order equations associated with large oscillations of the iterative solution
due to very large correlations in the $4f$ shell. Both the Yb$^{2+}$ core and some of the Yb$^+$
valence shell all-order equations that are used to construct the effective Hamiltonian diverge
using conventional iteration schemes. We have resolved this problem by using RLE and DIIS
convergence stabilizers described in \cite{RLE} within the framework of the CI+all-order method.
The main idea of these approaches is to accumulate several iterations and determine a next best
solution based on all
 stored data. Convergence
was achieved for the $ns$, $np$, and $(n-1)d$ valence states with $n=6-9$.

We present  the energy levels obtained in the CI, CI+MBPT, and CI+all-order approximations and
compare them with the experimental values \cite{RalKraRea11} in Table~I of the supplementary
material~\cite{EPAPS}. At  the CI stage, the theoretical energy levels differ rather significantly
from the experimental energies, up to 19\% for the $6s6p$ states. Including the core-valence
correlations in the second order of the MBPT improves the agreement to the 1.5-5.5\% level.
Further improvement of the theoretical energies is achieved when the CI+all-order approximation is
used. The two-electron binding energy is accurate to 0.7\% with the \textit{ab initio} CI+all-order
approach, a factor of 2 improvement in comparison with the CI+MBPT result.

The valence part of the polarizability is determined  by solving the inhomogeneous equation of
perturbation theory in the valence space, which is approximated as
\begin{equation}
(E_v - H_{\textrm{eff}})|\Psi(v,M^{\prime})\rangle = D_{\mathrm{eff},q} |\Psi_0(v,J,M)\rangle
\label{e11}
\end{equation}
for a valence state  $v$ with the total angular momentum $J$ and projection $M$ \cite{KozPor99a}.
 The effective dipole operator
$D_{\textrm{eff}}$ includes random phase approximation (RPA)
corrections.
 The ionic core  part of the polarizability, $\alpha_c$, is
calculated separately in the RPA and is found to be $\alpha_c=6.4$~a.u. The small valence-core (vc)
$\alpha_{vc}$ term that corrects the ionic
 core polarizability for the presence of the valence
electrons is also calculated in the RPA; it is  equal to $-0.4$~a.u.
and $-0.2$~a.u. for the $6s^{2}~^1S_0$ and $6s6p~^3P_0$ states,
respectively.  DHF calculations are carried out as well for both of
these contributions to evaluate the uncertainty associated with these
terms, which was found to be negligible at the present level of
accuracy.

Accurate calculation of the polarizabilities of low-lying states
 is more difficult for Yb than for alkaline-earth atoms. It is known that the main
contribution to the ground state polarizability of Yb comes from $4f^{14} 6s6p \,\,^1\!P^o_1$ and
$4f^{13} 5d6s^2\, (7/2,5/2)^o_1$ states (see, e.g.,~\cite{ZhaDal07}). The energy difference between
these states is only 3790 cm$^{-1}$ and they strongly interact with each other.  Calculations that
treat Yb as an atom with only two valence electrons fail to  account properly for the interaction
between valence and core-excited states and describe states with an unfilled $f$ shell. While the
state $4f^{13} 5d6s^2\, (7/2,5/2)^o_1$ does not belong to the valence subspace and is not directly
mixed with the $4f^{14} 6s6p \,\,^1\!P^o_1$ state in our calculations,
 its effect is introduced  via the
calculation of the effective Hamiltonian, since we allow all single
and double excitations of the core shells during its construction. As
a result, the polarizability calculation carried out via the solution
of the inhomogeneous equation does not appear to be affected by this
problem. A theoretical explanation of this fact was suggested
in~\cite{DzuDer10} which considered mixed and unmixed basis sets that
included $4f^{14}6s6p\,\,^1\!P^o_1$
 and $4f^{13} 5d6s^2\, (7/2,5/2)^o_1 $ states.
Excellent agreement of our results with all measured  Yb polarizability-related properties,
including Stark shift and magic wavelength of the  $6s^2 \, ^1\!S_0 - 6s6p \,\,^3\!P^o_0$
transition and the $C_6$ coefficient of the Yb-Yb dimer, confirms that the mixing problem does not
appear to affect such properties. We note that this is only true as long as no experimental data is
substituted for theoretical quantities in any part of the calculations, since this will compromise
the basis set completeness~\cite{DzuDer10}. It follows that  the direct solution of the
inhomogeneous equation is expected to be more accurate than expected based on comparison of
individual matrix elements with experiment.   This conclusion is important for future calculation
of the parity-violating amplitudes that could by evaluated by the same techniques.

Table~\ref{alpha_l} presents results for the static polarizabilities of the $6s^2 \, ^1\!S_0$ and
$6s6p \,\,^3\!P^o_0$ states and their differences. We note that the states with an unfilled $4f$
shell contributed less to the polarizability of the $^3\!P^o_0$ than $^1S_0$ state. In particular,
even-parity states with an unfilled $4f$ shell lie rather high in energy and their contributions to
the polarizability and influence on other states is not so significant.
\begin{table}
\caption{The $6s^2 \, ^1\!S_0$ and $6s6p \, ^3\!P^o_0$ static
polarizabilities $\alpha_g(\omega=0)$ of Yb  and their difference
%$ \Delta\alpha(^3\!P_0)-(^1\!S_0)$
 $\Delta\alpha \equiv \alpha(^3\!P_0^o) -\alpha(^1\!S_0)$
calculated in CI, CI+MBPT, and CI+all-order approximations in a.u.
The CI+all-order values are taken as final. The present results are
compared with other theoretical and experimental values.}
 \label{alpha_l}
\begin{ruledtabular}
\begin{tabular}{lccc}
%Method & $\alpha_0(6s^{2}~^1S_0)$ &  $\alpha_0(6s6p \, ^3\!P^o_0)$ &$ \Delta\alpha(^3P_0-^1S_0)$\\
Method & $\alpha(^1\!S_0)$ &  $\alpha(^3\!P_0^o)$ &$ \Delta\alpha$\\
\hline
CI & 187.9 & 279.7&\\
CI+MBPT &  138.3 & 305.9 &\\
CI+all-order &  140.9 &  293.2 &\\
Final & 141(3) & 293(10) & 152 \\
\hline
Theory~\cite{PorDer06} (2006)&  111.3(5) & 266(15)&155\\
Theory~\cite{ZhaDal07} (2007)& 143& & \\
Theory~\cite{SahDas08} (2008)& 144.6 &&\\
Theory~\cite{DzuDer10} (2010) &141(6) & 302(14)&161\\
Ref.~\cite{Bel12}$^{a}$ (2012) & 134.4 $-$ 144.2 & 280 $-$ 290& \\
Expt. \cite{YbclockBBR} (2012) &&& 145.726(3)
\end{tabular}
\end{ruledtabular}
$^{a}$Constraints based on experimental data. The uncertainty in each of these values is 1.0.
\end{table}
The results obtained in the CI, CI+MBPT+RPA, and CI+all-order+RPA approximations are presented. Our
recommended values obtained at the CI+all-order+RPA stage are in a reasonable agreement with other
theoretical values.  We emphasize that our calculations are completely \textit{ab initio}. The most
recent recommended values of Ref.~\cite{DzuDer10} include adjustment to reproduce the experimental
value of the magic wavelength.  The set of accurate experimental data  was used to set upper and
lower bounds on the $^1\!S_0$ and  $^3\!P^o_0$ polarizabilities in ~\cite{Bel12}. Our recommended
values are in excellent agreement with these constraints taking an account the uncertainties. We
can roughly estimate the uncertainty of our calculations as the difference of the CI+MBPT and
CI+all-order values, which yields 1.8\% and 4.3\%  for $^1S_0$ and $^3P_0^o$ states. We note that
the CI+all-order value is higher than CI+MBPT for $^1S_0$ but lower for $^3P_0^o$, so we can expect
that these uncertainties will add cumulatively for the $\Delta\alpha$ polarizability difference.
However, our value of $\Delta\alpha$ agrees with a recent experiment to 4.3\%, so our values are
somewhat more accurate than the estimates above ( 1\% and 3.5\%, respectively). A direct
measurement of the ground state polarizability with 1\% accuracy would be an excellent test of the
quality of calculations.

To further check the accuracy of our approach we calculated the magic
wavelength  $\lambda$ for the
$^1\!S_0$ and  $^3\!P^o_0$ states.  At the magic wavelength, the frequency-dependent polarizabilities
of the two states are equal.
We obtain $\lambda = 754$ nm  in the CI+all-order approximation which is
 within 1\% of the experimental value
759.355 nm~\cite{BarStaLem08}. The polarizability of the $^3\!P^o_0$ state grows rapidly
in the vicinity of the intersection of the ac polarizabilities. It means that even a small change in
$\lambda$ leads to a significant change in
$\alpha(^3\!P^o_0)$. For example,  the CI+MBPT value is significantly higher, 789~nm. Such close
agreement of the CI+all-order value
with the experimental wavelength confirms the
accuracy of the polarizabilities quoted above.

An important application of the polarizability calculation is to
determine the shift of the $^1\!S_0 - \,^3\!P_0^o$ transition
frequency by the effects of the ambient blackbody radiation. The BBR
shift is now one of the largest irreducible contributions  to the
budget of the uncertainty of optical atomic clocks. The leading
contribution to the  BBR shift of the energy level $g$ can be
expressed in terms of its static polarizability $\alpha_g(\omega=0) $
by \cite{PorDer06BBR}
\begin{eqnarray}
\delta E_{g} = -\frac{2}{15} (\alpha \pi)^3 (k_B T)^4
\alpha_g(0) \left[ 1 + \eta \right] ,
\label{delEg}
\end{eqnarray}
where $k_B$ is the Boltzmann constant, $T$ is the temperature, and
$\eta$ is  a ``dynamic'' fractional correction to the total shift
that reflects the averaging of the frequency dependence
polarizability over the frequency of the blackbody radiation
spectrum.

When the parameter $|y_n|>10$, where  $y_n=(E_n-E_g)/(k_B T)$,
$\eta$ can be  be approximated by ~\cite{PorDer06BBR}
\begin{eqnarray}
\eta &=& \eta_1+\eta_2+\eta_3=\frac{80}{63\,(2J_g+1)} \, \frac{\pi^2}{\alpha_g(0) k_B T} \nonumber \\
&& \times  \sum_n \frac{ |\langle n||D||g \rangle|^2 }{y_n^3} \left(
1+ \frac{21 \pi^2}{5\,y_n^2} + \frac{336 \pi^4}{11y_n^4} \right) ,
\label{eta}
\end{eqnarray}
with 0.1\% accuracy.
\begin{table}
\caption{The values of the $C_6$ coefficients (in a.u.) for the homonuclear Yb dimer and the
heteronuclear alkali-metal/group II - Yb dimers. All atoms are in their ground states. The
$\alpha(i\omega)$ for alkali and Mg, Sr, and Ca are taken from Ref.~\cite{DerPorBab10} in rows
CI+MBPT and  CI+all$^{(a)}$. The
 $\alpha(i\omega)$ for Mg, Sr, and Ca are calculated with the
 CI+all-order method in the present work in row CI+all$^{(b)}$.
  The present CI+MBPT and CI+all-order Yb $\alpha(i\omega)$ values are used in rows
    CI+MBPT and CI+all-order$^{(a,b)}$, respectively. The uncertainty
     of the final CI+all-order
    values is estimated to be 2\%. } \label{C6_l}
\begin{ruledtabular}
\begin{tabular}{lccccl}
                              &   Li-Yb&Na-Yb&K-Yb &  Rb-Yb&Cs-Yb  \\
\hline
 CI+MBPT      &  1534 & 1655 & 2548 & 2807 & 3367  \\
 CI+all$^{(a)}$& 1551 & 1672 & 2576 & 2837 & 3403 \\
 \hline
           &Yb-Yb &Mg-Yb&Ca-Yb&Sr-Yb&   \\
   \hline
    CI+MBPT      &  1901& 1086 &2000&2414&  \\
    CI+all$^{(a)}$&  &1093 &2017&2435&  \\
    CI+all$^{(b)}$&  1929          &2024&    &      \\
   Expt.~\cite{KitEnoKas08}  &  1932(35)          &  &  \\
\end{tabular}
\end{ruledtabular}
\end{table}
We express the dominant term in the equation above as the second
derivative of the polarizability:
\begin{equation}
\eta_1 \approx \frac{20}{21\,(2J_g+1)} \, \frac{(\pi k_B
T)^2}{\alpha_g(0)}
%\frac{\partial^2}{\partial E_g^2}\alpha_{+}^{\omega}.
\frac{\partial^2}{\partial E_g^2}\alpha_{g}(0)
\end{equation}
and find $\eta_1(^1S_0)=0.00116$ and $\eta_1(^3P_0)=0.00934$.
 We calculated the second term in Eq.~(\ref{eta}) using both
a forth derivative of $\alpha$ and sum over states with the CI+all-order values of the matrix
elements; identical result $\eta_2(^3P_0)=0.00029$ was obtained. $\eta_2$ is negligible for
$^1S_0$, 0.000003.
 The third term can be neglected at the present level of accuracy.
 The resulting values of the dynamic corrections to the BBR shift at
 300~$K$ are $\delta\nu_{\rm BBR}(^1\!S_0)=-0.0014$~Hz and
 $\delta\nu_{\rm BBR}(^3\!P_0)=-0.0243$~Hz, respectively.

The  total dynamic correction to the BBR shift at 300K is determined
as the difference between the individual shifts, $\delta\nu_{\rm
BBR}^{\rm dyn} = -0.0229$~Hz. Combining this result with the
experimental determination of the  $\Delta \alpha=145.726(3)$~a.u.
\cite{YbclockBBR}, we get the final result for the BBR shift at
300~$K$:
 $\delta\nu_{\rm BBR} = -1.2777(8)$~Hz.
This value is in excellent agrement with the determination of the BBR shift mostly from the
experimental data  $\delta\nu_{\rm BBR} = -1.2774(6)$~Hz that was just reported in ~\cite{new}.
 We have verified that the M1 and E2
 contributions to the BBR shift are negligible at the present level of accuracy.
 Details of the calculation of the
 dynamic correction to the BBR shift are given in the supplementary
 material~\cite{EPAPS}.

 Many of the same considerations concerning accurate calculation of
the frequency-dependent polarizability arise in the calculation of the van der Waals coefficients.
If two atoms $A$ and $B$ have spherically symmetrical ground states, the leading power of the
long-range interactions takes the form $ V(R)=- C^{AB}_{6}/R^{6}, \label{VRn} $ where $R$ is the
distance between atomic nuclei. The the van der Waals coefficient $C^{AB}_{6}$  can be calculated
as \cite{PatTan97}
\begin{equation}
C^{AB}_6 = \frac{3}{\pi}\, \int_0^\infty\, \alpha^A(i \omega)\,
\alpha^B(i \omega) d\omega,
\label{C6}
\end{equation}
where $\alpha(i \omega)$ is the frequency-dependent polarizability at an imaginary frequency. In
practice, we compute the $C^{AB}_6$ coefficients  by approximating the integral (\ref{C6}) by
Gaussian quadrature of the integrand computed on the finite grid of discrete imaginary frequencies
~\cite{BisPip92}.

For the alkali and group II  atoms, we use frequencies and weights tabulated in
Ref.~\cite{DerPorBab10} at 50 points. These dynamic polarizabilities were obtained by combining
high-precision experimental data for matrix elements of principal transitions with high-precision
many-body methods, such as linearized coupled-cluster approach and CI+MBPT.  The accuracy of the
corresponding homonucelar $C_6$ was estimated to be better than 1\% for all cases relevant in this
work with the exception of Ca, were it was 1.5\% ~\cite{DerPorBab10}.

 The Yb imaginary frequency polarizabilities $\alpha_(i\omega)$
 for the ground $^1\!S_0$ state are calculated in this work by solving
the inhomogeneous equation (\ref{e11}) with the appropriate modifications. We use the same 50-point
frequency grid as in Ref.~\cite{DerPorBab10} for consistency. To evaluate the uncertainty in the
$C_6$ coefficients, we carried out both CI+MBPT and CI+all-order calculations of the ground state
$\alpha(i\omega)$. The same alkali and group II data are used in both cases. The results are
summarized in Table~\ref{C6_l}. We find that the differences between CI+MBPT and CI+all-order
results are actually smaller (1-1.5\%) than for the ground state static polarizability (1.8\%)
since the differences decrease with $\omega$ for $\alpha(i\omega)$. As a result, we expect the
accuracy of the $C_6$ to be on the same order as the static polarizability, rather than larger by a
factor of two. Moreover, our value for the $C_6$ coefficient of the homonuclear Yb dimer is in
excellent agreement with the experimental result ~\cite{KitEnoKas08}, which is accurate to 1.8\%.
Based on the comparison of the CI+MBPT and CI+all-order values for heteronuclear $C_6$
coefficients, and agreement with experiment for the Yb $^3P_0-^1S_0$ Stark shift and  magic
wavelengths, and $C_6$ coefficient for Yb-Yb dimer, we estimate that our predictions of the $C_6$
coefficients for the heteronuclear alkali-metal atom/group II - Yb dimers are accurate to about
2\%.

  In conclusion, we have carried out fully \textit{ab initio}
  all-order calculations of Yb properties. Our values of the Yb
$^1\!S_0-\,^3\!P_0^o$ Stark shift and magic wavelength as well as the $C_6$ coefficient of the
Yb$_2$ dimer are in excellent agreement with experiment. We have developed a new approach of
calculation of the dynamic correction to the BBR shift that does not involve an explicit sum over
states. The Stark shift of the clock transition was determined experimentally~\cite{YbclockBBR}
with a high precision. As a result, the uncertainty in the dynamic correction can now be directly
related to the uncertainty of the BBR shift of this transition. Thus, when combined with the recent
measurement of the Yb clock Stark shift ~\cite{YbclockBBR}, our calculation of the dynamic
correction allows us to reduce the fractional uncertainty due to the BBR shift in the Yb optical
lattice clock to $10^{-18}$ level. The same method can be used to evaluate the dynamic correction
for any optical atomic clock. Finally, we have presented the first recommended values of $C_6$
coefficients for  alkali/group II-Yb dimers for future experimental efforts in producing ultracold
polar molecules.

We thank P. Julienne, C. Oates, and J. Sherman for helpful
discussions. This research was performed under the sponsorship of the
U.S. Department of Commerce, National Institute of Standards and
Technology, and was supported by the National Science Foundation
under Physics Frontiers Center Grant No. PHY-0822671 and by the
Office of Naval Research. The work of S.G.P. was supported in part by
US NSF
 Grant No.\ PHY-1068699 and RFBR Grant No.\ 11-02-00943.
%\bibliography{YbC6}

\end{document}